\documentclass[aip,apl,reprint]{revtex4-1}

\usepackage{amsmath}
\usepackage{graphicx}
\usepackage{pstricks}
\usepackage{upgreek}

\newcommand{\unit}[1]{\,\text{#1}}

\newcommand{\outline}[1]{}

\newcommand{\sqr}[1]{\left[ #1 \right]}

\begin{document}

\title{Ordering of block copolymer microstructures in corner geometries}

\author{W. T. Lee} \homepage{http://www.ul.ie/wlee}
\email{william.lee@ul.ie} \author{N. Delaney} \author{M. Vynnycky}
\affiliation{MACSI, Department of Mathematics and Statistics,
  University of Limerick, Limerick, Ireland.}

\author{M. A. Morris} \affiliation{Department of Chemistry, University
  College Cork, Cork, Ireland.} \affiliation{Centre for Research on Adaptive
  Nanostructures and Nanodevices (CRANN), Trinity College Dublin,
  Dublin, Ireland.}



\begin{abstract}
The ordering of block copolymers into lamellar microstructures is an
attractive route for creating nanopatterns on scales too small to be
constructed by current photolithography techniques. This utilises a
technique known as graphoepitaxy where topography is used to define
the alignment of the pattern for precise placement of pattern
features.  One problem with this approach is the failure of lamellae
to maintain continuity around corners, due to geometrical
frustration. We report simulation results using the extended
Cahn-Hilliard equation which suggest that this problem could be solved
by using rounded corners.
\end{abstract}


\maketitle

\section{Introduction}

The self assembly of block copolymers is exploited in many areas of
nanotechnology, e.g.~energy storage and conversion\cite{Orilall2011}
and drug delivery\cite{Nostrum2011}. In this work we focus on the use
of block copolymers to template nanopatterns at substrate surfaces on
scales too small to be constructed by traditional top down
photolithography.\cite{Farrell2009,Farrell2010,Fitzgerald2009} The
patterns formed have to be ‘directed’ using chemical pre-patterning or
topography.  This topographical guiding of the pattern is known as
graphoepitaxy and can provide excellent pattern
registry.\cite{Stuen2010} One problem with this form of directed
self-assembly is the difficulty in defining complex features such as
curves, bends and junctions.  This work centres on the particular
issue of defect-free pattern alignment persisting around
corners.\cite{ESGI62} We use simulations to investigate whether this
problem could be overcome by using rounded corners of the guiding
features.

Block copolymers are polymers consisting of blocks of different monomer
units. The simplest such form is the diblock copolymer which consists
of two different monomers `a' and `b' joined by a single covalent
bond. Schematically the structure of the monomer is 
\begin{center}
\ldots-a-a-a-a-a-a-a-b-b-b-b-b-b-b-\ldots
\end{center}
which can be simplified to A-B, where A and B represent chains of a
and b monomers respectively. These polymers show interesting behaviour
when the monomers a and b have distinct chemical differences, for
instance if a is hydrophillic while b is hydrophobic. In this case a
and b monomer units will be attracted to other monomer units of the
same type while being repelled by the other type. Thus the most
energetically stable state of the system, i.e.\ the state that will be
stable below some order-disorder phase transition temperature, is one
in which the system is segregated into regions of a monomer units and
regions of b monomer units.

The structure of the polymer molecules constrains the degree to which
this segregation can occur. The a and b monomer units are part of the
same molecule. Regions of a and b can never be separated by a distance
larger than the size of single polymer molecule. In the ordering of
block copolymers the order parameter is locally conserved over a
region the size of a single molecule, unlike conventional ordering
processes where the order parameter is globally conserved but there is
no limit on the degree of coarsening of the microstructure, except
that imposed by the kinetics of ordering.

\begin{figure}
\begin{center}
\includegraphics{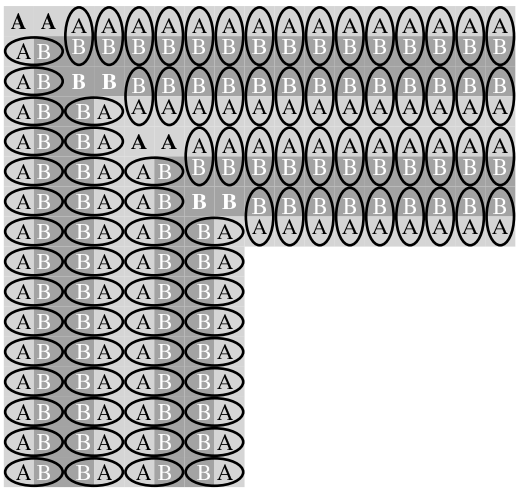}
\end{center}
\caption{\label{geometric_frustration} In this figure A represents a
  chain of -a- monomers, B represents a chain of -b- monomers, thus
  the formula of a diblock copolymer is A-B. As the figure shows a
  corner geometry cannot be constructed from A-B units. }
\end{figure}

As detailed above, surface topography is used to direct the block
copolymer pattern to create ordered alignment.  In the simplest form
of graphoepitaxy, conventional lithographic techniques are used to
create trenches in a silicon wafer. The trenches are filled with block
copolymer which orders into lamellae parallel to the sidewall on
annealing.\cite{Farrell2009,Farrell2010,Fitzgerald2009,Stuen2010} The
lamellae are on a much finer scale than the trench. Finally a selected
monomer is chemically etched away and the remaining polymer used as an
etch mask to facilitate pattern transfer to the substrate, creating
nanowires on a scale too fine to be manufactured by conventional
lithographic techniques.\cite{Morris2012,Borah2011} The failure of the
pattern and hence the nanowires to persist around corners can be
attributed to geometric frustration.\cite{ESGI62} The ordering pattern
required to maintain continuity around a corner is inconsistent with
the condition that the order parameter must be locally conserved, as
illustrated in Figure~\ref{geometric_frustration}.

In this work, we use simulations to investigate whether this problem
can be overcome by using rounded corners. Simulations on the unit disk
show that ordering patterns of concentric circles form.\cite{Ren2002}
Thus the key remaining question is whether ordering patterns in the
rounded corners are compatible with patterns in the straight trenches.
We compare the topology (number of lamellae) of the microstructure in
straight and annular geometries and show that once the inner radius
exceeds a critical value the microstructures have the same topology
suggesting that, in a rounded corner geometry, nanowires will persist
around corners.

\section{Methods}

To simulate the block copolymer ordering we use the extended
Cahn-Hilliard equation\cite{Choksi2003,Wu2006}, which can be written in
dimensionless form as
\begin{equation}
\dfrac{\partial\phi}{\partial t}
         =\nabla^2 \sqr{ -\phi + \phi^3 -\nabla^2 \phi } -\alpha\phi. 
\end{equation}

\begin{figure}
\begin{center}
\includegraphics[]{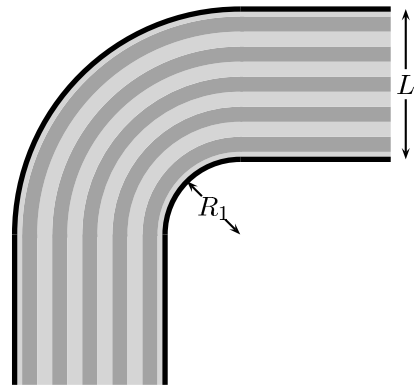}
\end{center}
\caption{\label{real_geometry} Rounded corner geometry. The inner
  radius of the corner is $R_1$ and the width of the trench is
  $L$. This geometry is difficult to simulate directly if the aim is
  to explore a large area of phase space. }
\end{figure}

\begin{figure}
\begin{center}
\includegraphics{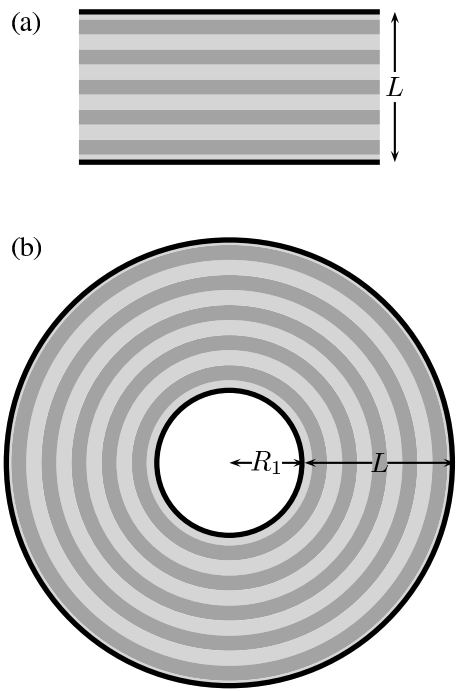}
\end{center}
\caption{\label{simulated_geometry}In this work we simulate ordering
  in two geometries. (a) An infinite straight trench. (b) An annular
  ring. The simplicity of these geometries allows us to carry out one
  dimensional simulations allowing a large region of parameter space
  to be investigated. }
\end{figure}

The geometry we are interested in is illustrated in
Figure~\ref{real_geometry}.  However we carry out simulations in two
separate geometries, illustrated in
Figure~\ref{simulated_geometry}. These are a straight region and an
annular ring. These geometries allow one dimensional simulations to be
carried out, making it possible to efficiently survey large regions of
parameter space.  We argue that, if solutions of the same topology
exist in both the straight region and the annular ring, this provides
strong evidence that lamellae will persist around rounded corners with
the same parameters. We choose to simulate the case in which there are
five lamellae in the straight trench, a situation realised
experimentally, for example, in Ref.~\onlinecite{Farrell2010}, for a
trench $220\unit{nm}$ wide.

Thus the equations we are solving are
\begin{equation}
\dfrac{\partial\phi}{\partial t}
         =\dfrac{\partial^2}{\partial x^2} 
                  \sqr{ -\phi + \phi^3 -\dfrac{\partial^2 \phi}{\partial x^2} } -\alpha\phi 
\end{equation}
with boundary conditions
\begin{equation}
\dfrac{\partial \phi}{\partial x}=\dfrac{\partial^3 \phi}{\partial x^3}=0
\end{equation}
at $x=0$ and $x=L$ in the straight geometry and
\begin{equation}
\dfrac{\partial\phi}{\partial t}
  =\dfrac{1}{r}\dfrac{\partial}{\partial r} r \dfrac{\partial}{\partial r}  
  \sqr{ -\phi + \phi^3 
  -\dfrac{1}{r}\dfrac{\partial}{\partial r} r \dfrac{\partial \phi}{\partial r}
           } -\alpha\phi 
\end{equation}
with boundary conditions
\begin{equation}
\dfrac{\partial \phi}{\partial r}
=\dfrac{\partial}{\partial r}\dfrac{1}{r}
     \dfrac{\partial}{\partial r} r \dfrac{\partial\phi}{\partial r}=0
\end{equation}
at $r=R_1$ and $r=R_2=R_1+L$ in the annular geometry.

Numerical simulations were carried out using the finite difference
method on a grid of 400 points. Timesteps of 0.1 dimensionless units
were carried out using the implicit Euler method. We took $\alpha=0.1$
following Ref.~\onlinecite{Wu2006}. We start the straight simulations
with a small amplitude cosine wave with 4.5, 5 and 5.5 wavelengths in
the domain. We determined the largest $L$ values at which these
solutions are stable by the bisection method\cite{NR}. 
This allows us to determine the values of $L$  in which the longest
wavelength solution to the extended Cahn-Hilliard equation has five
lamellae (in a straight geometry). 

We then sample $R_1$ and $R_2=R_1+L$ space. Again we start simulations
with a small amplitude cosine wave, with five wavelengths in the
domain, and the region of interest is one in which the final solution
also has five wavelengths, i.e. five lamellae. As before we use
bisection to find the limits of stability of this solution. This
allows us to construct a `phase diagram' showing regions in which
solutions to the Cahn-Hilliard equation have the same topology in
straight and annular geometries.

As a final check we carried out a two dimensional finite element
simulation on a rounded corner geometry using the COMSOL
software. The simulation geometry has $L=70$ and $R_1=75$.

\section{Results}

\begin{figure}
\begin{center}
\includegraphics{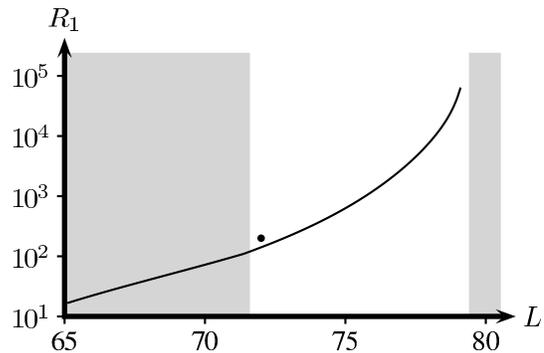}
\end{center}
\caption{\label{phase_diagram} `Phase diagram'. Above the line shown
  five wavelength solutions to the extended Cahn-Hilliard equation are
  stable. The region shown in white (as opposed to grey) is the values
  of $L$ for which the five wavelength solution is the longest
  wavelength stable solution in a straight geometry. The black circle
  shows the parameters used in the simulation shown in
  Figure~\ref{compare}.}
\end{figure}

\begin{figure}
\begin{center}
\includegraphics{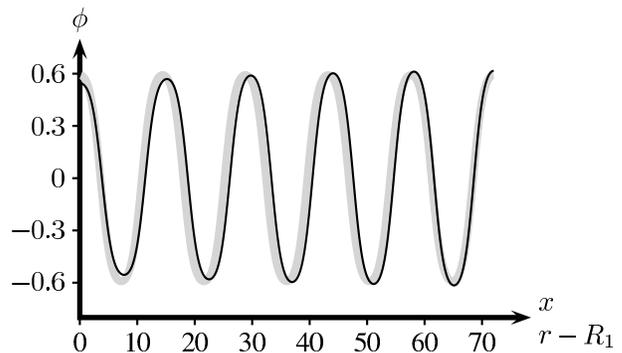}
\end{center}
\caption{\label{compare} Comparison of solutions to the Extended
  Cahn-Hilliard equation in straight (grey line) and annular (black
  line) trenches. The graph shows the solutions have the same topology
  suggesting compatibility of lamalae in a straight trench and a
  rounded corner.}
\end{figure}

\begin{figure}
\begin{center}
\includegraphics[]{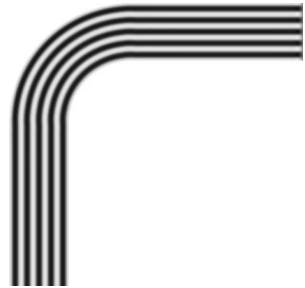}
\end{center}
\caption{\label{two_dimensional}Finite element simulation of a corner
  geometry with $L=70$ and $R_1=75$. }
\end{figure}

The one dimensional simulation results are summarised in
Figure~\ref{phase_diagram}. They show that solutions with five
lamellae are the longest wavelength stable solutions in straight
geometries with $71.6<L<79.4$. Solutions with the same topology occur
in annular geometries for long enough wavelengths as shown in the
`phase diagram' in Figure~\ref{phase_diagram}. Five wavelength
solutions to the extended Cahn-Hilliard equation in straight and
annular geometries are shown in Figure~\ref{compare}. The solutions
are reasonably similar, with the largest discrepancy occurring close to
the origin, where the curvature is largest. The results of the two
dimensional simulation is shown in Figure~\ref{two_dimensional} and
confirms that lamellae can persist round corners.

\section{Discussion}

The coexistence of solutions to the Cahn-Hilliard equations with the
same topology in straight and annular geometries of the same width
suggests that rounded corners could be a solution to the problem of
templating or pattern transferring nanowires in complex geometries. 

\begin{figure}
\begin{center}
\includegraphics{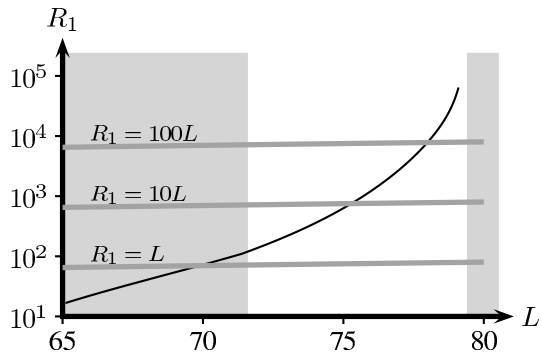}
\end{center}
\caption{\label{phase_diagram2} `Phase diagram' from
  Fig.~\ref{phase_diagram}. The gray lines show the curves $R_1=L$,
  $R_1=10L$, and $R_1=100L$. Solutions of five wavelengths coexist in
  rounded and straight geometries for $R_1\lesssim 10L$ and
  $R_1\lesssim100L$.}
\end{figure}

One important remaining question, addressed in
Figure~\ref{phase_diagram2}, is whether this coexistence occurs in a
useful region of parameter space. If the values of $R_1$ at which
coexistence occur are too large it will make the technique
impractical, since the rounded corners will occupy too much space on
the die. The figure shows that the ideal case of coexistence for
$R_1\approx L$ does not occur. However, there is a region of phase
space for which coexistence occurs for $R_1<10L$. This still seems
small enough to make the technique potentially practical. A still
larger region of parameter space is available if $R_1<100L$ is
allowed, but this may be too large to make the technique
usable. Taking the trench width as $L=220\unit{nm}$ (from
Ref.~\onlinecite{Farrell2010}), then $10L=2.2\unit{}\upmu\text{m}$ and
$100L=22\unit{}\upmu\text{m}$. At least the first of these seems small
enough to be useful in devices. The situation modelled in
Figure~\ref{compare} would correspond to $R_1\approx 600\unit{nm}$
suggesting that sub-micron values of $R_1$ may be practical.  As
Figure~\ref{phase_diagram2} shows, to minimise $R_1$, the $L$ should
be made small as possible given the desired number of lamellae.

\section{Conclusions}

The ordering of block copolymers into lamellar microstructures offers
an attractive route to pattern transfer or template nanowires on a
scale too small to be created by conventional lithographic
techniques. One problem with this approach is the difficulty in
maintaining continuity of the nanowire templates in complex
geometries, such as corners. We have used simulations to investigate
the feasibility of using rounded corners to overcome this problem. Our
results suggest that this approach will work if the inner radius of
curvature of the rounded corner is large enough. Rough estimates
suggest that for a straight trench about $0.2\unit{}\upmu\text{m}$
wide containing five lamellae lamellae will persist around a rounded
corner with a radius less than $2\unit{}\upmu\text{m}$.

\begin{acknowledgments}
WTL, ND and MV acknowledge support of the Mathematics Applications
Consortium for Science and Industry (\url{http://www.macsi.ul.ie})
funded by the Science Foundation Ireland Mathematics Initiative Grant
06/MI/005.
This work is part supported by the EU project LAMAND (contract Nr.
245565). The contents of this work are the sole responsibility of the
authors.
The support of the CRANN SFI CSET project is also
acknowledged.
\end{acknowledgments}

%

\end{document}